\documentclass[12pt]{iopart}
\usepackage{graphicx} % standard LaTeX graphics tool
\begin{document}
\title{Propagation of a mutually incoherent optical vortex pair in 
anisotropic nonlinear media}
\author{A. V. Mamaev\dag, M. Saffman\ddag, and A. A. Zozulya\S}
\address{\dag Institute for Problems in Mechanics, Russian Academy of Sciences, 
Prospekt Vernadskogo 101, Moscow, 117526 Russia } 
\address{\ddag Department of Physics,
University of Wisconsin, 1150 University Avenue,  Madison, Wisconsin 53706, USA } 
\address{\S Department of Physics, WPI, 100 Institute Road, Worcester, Massachusetts 01609, USA}
\begin{abstract}
We study propagation of a pair of oppositely charged and mutually incoherent vortices 
in anisotropic nonlinear optical media. Mutual interactions retard the delocalization 
of the vortex core observed for isolated vortices.  
\end{abstract}
\date{\today}
\maketitle

\section{Introduction}
\label{introduction}
Isotropic media with a repulsive nonlinearity support localized vortex solitons 
characterized by a dislocation of the wavefront at the point where the field 
amplitude vanishes \cite{GinPit}. 
Vortex solitons have been observed in several types of defocusing nonlinear media, including 
superfluids, superconductors, optical Kerr media  and Bose Einstein condensates. 
Stability of the solitons 
depends on the structure of the nonlinearity and the dimensionality of the 
medium.  Planar one-transverse dimensional [(1+1)D] solutions are modulationally unstable in bulk,
two-transverse dimensional [(2+1)D] 
media \cite{ZakRub,stripe_exp,stripe_exp2}. Kerr type optical media with a cubic, 
isotropic, and local nonlinearity support stable (2+1)D vortex solitons \cite{GinPit,swa}. 
These vortex solitons have attracted considerable attention theoretically\cite{nlovortex} and have recently been studied in detail in atomic condensates\cite{atomicvortex}. 
A Kerr type nonlinearity is, however, a simplified idealized model of a nonlinear response. 
The bulk photorefractive nonlinear medium used in the experiments reported here 
exhibits a nonlinearity that is both anisotropic and nonlocal. This leads to 
qualitative differences in the spatial dynamics of vortex beams. 

One of the most basic phenomena is the propagation of an isolated vortex of unit 
topological charge. Such a vortex can form a stable soliton in isotropic 
defocusing media\cite{swa}. Theory and experiments have shown that unit charged vortices become delocalized
when they propagate in photorefractive media\cite{vrpair,nocircular,vrtransient}. 
The nonlinearity in a photorefractive crystal with an externally applied electric field, or photogalvanic 
response, is anisotropic 
and nonlocal\cite{ZAPRA}. The nonlinearity induced by a localized, circular beam is roughly 3 times stronger along the 
direction of the applied field than in the perpendicular direction (we will refer to the direction of strongest nonlinearity as the $\hat z$ axis).  Due to the anisotropy a vortex with an initially azimuthally symmetric core profile focuses 
perpendicular to $\hat z,$ and stretches along $\hat z$, 
so that the major axis of the core coincides with the direction of greatest material nonlinearity.  
Simultaneously the elongated vortex starts to rotate due to its phase structure. The direction of rotation is uniquely 
determined by the vortex charge; changing its sign changes the sign of the rotation. 
Eventually the rotation is stopped by the anisotropy so that the major axis of the vortex 
is aligned at some angle with respect to $\hat z.$  
The stretching, however,  proceeds unchecked so that the vortex becomes more and more delocalized. 
Both theory and experiment show that delocalization of the vortex core is generic in anisotropic media, and not dependent on 
a specific choice of parameters. The implication is that localized optical vortex solutions and, in particular, 
vortex solitons of unit topological charge do not exist in these media. 

An additional hallmark of anisotropy is the nonlinear decay of a charge $n$ vortex into $n$ unit charge vortices. 
The decay, although expected on energetic grounds, does not occur in isotropic media where a high charge vortex 
is metastable\cite{ara96}, but was observed using a photorefractive crystal as the nonlinear medium\cite{vrdecay}. 
 Upon breakup of the input 
high-charge vortex the resulting charge-one vortices repel each other and form an array  aligned 
perpendicular to $\hat z$. The decay is driven by the intrinsic 
anisotropy of the medium and takes place for any core profile of the input vortex field.   
The theory developed in Ref. \cite{vrdecay} suggests that the decay of high-charge vortices 
is possible in local isotropic media provided some anisotropy is introduced in the problem via, e.g., 
initial boundary conditions.  
   
Given the instability of unit and high charge vortices in anisotropic media it is natural to ask if 
there exist self-bound field configurations that remain localized under propagation in anisotropic media. 
Previous experiments\cite{vrpair} have demonstrated that the effects of anisotropy are 
considerably weakened for a counterrotating pair of vortices with zero net topological charge.   
When such a pair is aligned perpendicular to $\hat z,$ 
it forms a bound state that translates parallel to $\hat z,$  
in a manner that is similar to the translational motion of 
a counterrotating point vortex pair in  fluid dynamics\cite{lamb}. 
The translational motion may, however, be inconvenient in the context of optical processing 
applications. Furthermore, the initial 
orientation of the pair is crucial for its subsequent evolution. A pair aligned along $\hat z$
annihilates due to diffraction.

In this paper we propose, and  demonstrate experimentally a novel technique of creating 
bound vortex structures in anisotropic media that are free of the above mentioned limitations. 
The main idea is based on the fact that in the prevailing majority of nonlinear media 
the material response is slow as compared to the frequency of the optical field, and therefore is a 
function of the time-averaged intensity of this field. If the light field consists of several 
features separated by frequencies that are fast on the time scale of the material response,  
the material response will be a function of the sum of these features intensities, whereas 
the cross-terms oscillating at high frequencies can be neglected. This fact has been known 
in photorefractive nonlinear optics and used successfully for implementing several optical 
processing devices \cite{toys}, as well as (1+1)D solitary structures \cite{incohsol}. 

We propose to create a bound vortex pair consisting of two copropagating counterrotating 
vortices sitting on top of each other that are mutually incoherent at the time scales of the 
nonlinearity. This incoherence can be achieved either by separating the carrier frequencies 
of the vortices by an amount that is larger than the inverse relaxation time of the medium, or 
profiling time histories of the input vortex fields such that their time overlap integrated 
over the relaxation time of the medium is zero. The mutual incoherence of the vortex fields 
removes the translational motion due to the linear coherence of the vortex pair. 
At the same time each of the vortices individually tries to rotate in opposite directions which 
effectively controls the anisotropy induced stretching of a single vortex.

\begin{figure}[!t]
\centering
\begin{minipage}[c]{7.5cm}
\includegraphics[width=7.5cm]{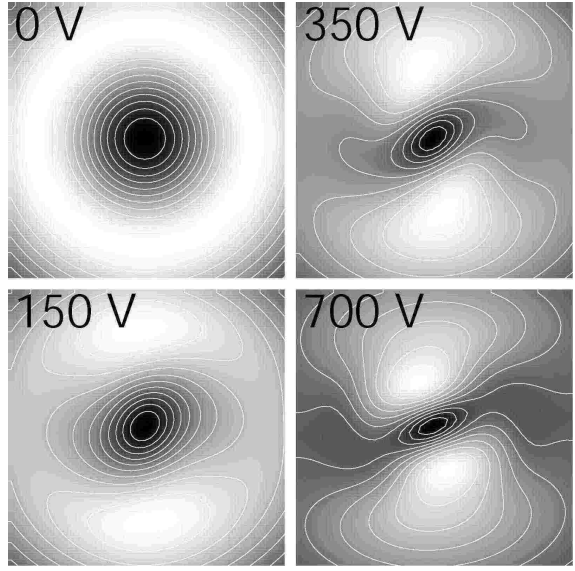}
\caption{Numerical results showing output spatial intensity distributions 
of a single vortex for different applied voltages.}
\label{fig.1}
\end{minipage}
\end{figure}

\section{Theory}
\label{theory}
In the theoretical analysis we use the set of equations describing 
propagation of an electromagnetic field $B(\vec{r})$ in a photorefractive self-focusing or 
self-defocusing medium as developed in \cite{ZAPRA,msazpra}. In our case 
the field consists of two temporal features. For definiteness assume that these features 
are separated by a frequency shift $\Omega$ such that $\Omega \tau \gg 1$, 
where $\tau$ is the characteristic response time of the nonlinearity:
\begin{equation}
B(\vec{r},t) = \left[ B_{+}(\vec{r},t) + B_{-}(\vec{r},t)\exp(-i  \Omega t)\right] 
\exp(i k x - i \omega t) \;.
\label{fields}
\end{equation}  
In the steady state the equations governing spatial evolution of the amplitudes   
$B_{\pm}$ take the form
%
%%%%%%%%%%%%%%%Maxwell %%%%%%%%%%%%%%%%%%%%%%%%%%%
\begin{numparts}
\begin{eqnarray}
&&\left[\frac{\partial}{\partial x} - \frac{i}{2}\nabla^{2}\right]B_{+}(\vec{r}) = 
-i\: \frac{\partial \varphi}{\partial z} B_{+}(\vec{r}) \;, \\
&&\left[\frac{\partial}{\partial x} - \frac{i}{2}\nabla^{2}\right]B_{-}(\vec{r}) = 
-i\: \frac{\partial \varphi}{\partial z} B_{-}(\vec{r}) \;, \\
&&{\nabla}^{2} \varphi +  \nabla \varphi \cdot \nabla \ln (1 + |B_{+}|^{2} + |B_{-}|^{2})   
= \frac{\partial}{\partial z} \ln (1 + |B_{+}|^{2} + |B_{-}|^{2}) \; .
\end{eqnarray}
\label{Maxwell}
\end{numparts}
%%%%%%%%%%%%%%%%%%%%%%%%%%%%%%%%%%%%%%%%%%%%%%%%%
Here $\nabla = \hat{y} (\partial /\partial y) + \hat{z} (\partial / \partial z)$  
and $\varphi$ is the dimensionless electrostatic potential induced by the light with  
the boundary conditions $\nabla \varphi(\vec{r} \rightarrow \infty) \rightarrow 0$. 
The dimensionless coordinates $(x,y,z)$ are related 
to the physical coordinates $(x',y',z')$ by the expressions $x = \alpha x'$ and $(y,z) = 
\sqrt{k \alpha}(y',z')$, where $\alpha = (1/2)k n^{2}r_{\rm eff} E_{\rm ext}$. Here  
$k$ is the wave number of light in the medium, $n$ is the index of refraction, 
$r_{\rm eff}$ is the effective element of the electro-optic tensor,  and 
$E_{\rm ext}$ is the amplitude of the external field directed along 
the $z$ axis far from the beam. The normalized intensity $I = |B_{+}|^{2} + |B_{-}|^{2}$ is 
measured in units of saturation intensity $I_{d},$ so that the physical beam 
intensity is given by $I \times I_d.$ The minus sign on the right hand side of 
Eqs. (\ref{Maxwell}a,b) corresponds to a self-defocusing nonlinearity.

Numerical solutions of Eqs. (\ref{Maxwell}) describing nonlinear evolution of a single 
vortex ($B_{-} = 0$) for different values of the applied voltage (nonlinearity) 
are shown in Fig. 1. The numbers on the frames are the values of this voltage in volts. 
Superimposed on the images are the equal intensity contour lines visualizing 
distortions of the vortex core for different values of the nonlinearity. 
The size of the frames is about $200~\mu \rm m$ . 

The input field was taken to be 
%
%%%%%%%%%%%%%%%%%%%%%single%%%%%%%%%%%%%%%%%%%%%%%%%
\begin{equation}
B_{+}(x=0,\vec{r}) = \sqrt{I_{\rm in}} r \exp(-r^{2}/d_{G}^{2} + i\theta)
\label{single}
\end{equation}
where $\theta$ is the azimuthal angle, $r = \sqrt{y^{2} + z^{2}}$ and $d_{G}$ is the diameter 
of the Gaussian beam. The parameters of the calculation were the following: the wavelength 
$\lambda = 0.63~\mu \rm m$, the refractive index $n = 2.3$, the length of the nonlinear 
medium $l = 2 ~\rm cm$, the effective electrooptic coefficient $r_{\rm eff} = 130 ~\rm pm/V$, 
$d_G = 115~\mu{\rm m}$ ($135~\mu{\rm m}$ Full Width at Half Maximum) and 
$I_{\rm in} = 1$.  

The output intensity distribution in the absence of the nonlinearity is given by 
the frame labeled $0~\rm V$  
and corresponds to an annular ring having approximately 
$79~\mu \rm m$ and $267~\mu \rm m$  internal and external diameters, respectively. 
The frames corresponding to nonzero nonlinearity demonstrate focusing of the vortex 
along $\hat y$ and its stretching along $\hat z.$ 
Also clearly seen is the rotation 
and alignment of the vortex. This rotation is charge-dependent and changes sign 
if the charge of the vortex is changed from plus to minus one. The magnitude of all 
these effects is directly proportional to the value of the applied voltage. 

 Numerical solutions of Eqs. (\ref{Maxwell}) describing nonlinear evolution of the 
mutually incoherent vortex pair for different values of the applied voltage (nonlinearity) 
are shown in Fig. 2.  The input field was taken to be 
%%%%%%%%%%%%%%%%%%%%%pair%%%%%%%%%%%%%%%%%%%%%%%%%
\begin{equation}
B_{\pm}(x=0,\vec{r}) = \sqrt{I_{\rm in}} r \exp(-r^{2}/d_{G}^{2} \pm i\theta)
\label{pair}
\end{equation}
All parameters were the same as in the case of a single vortex (Fig. 1). 
The parameter $I_{\rm in}$ was again chosen such that the total maximum input 
intensity was equal to one (the maximum intensity of each of the constituents 
of the pair was 0.5).  Figure 2 demonstrates some considerable reduction in the 
magnitude of anisotropy effects as compared to the case of a single vortex (Fig.1).
In particular, the degree of ellipticity of the vortex core at high voltages is several 
times smaller in Fig. 2 than in Fig. 1. The horizontal intensity dip appearing at high 
intensities on the Gaussian beam and passing from left to right is several 
times smaller in Fig.2 than in Fig. 1.  Figure 2 also confirms that the vortex pair 
remains stationary and does not translate with respect to the Gaussian beam.

%
%Fig.2
%
\begin{figure}[!t]
\centering
\begin{minipage}[c]{7.5cm}
\includegraphics[width=7.5cm]{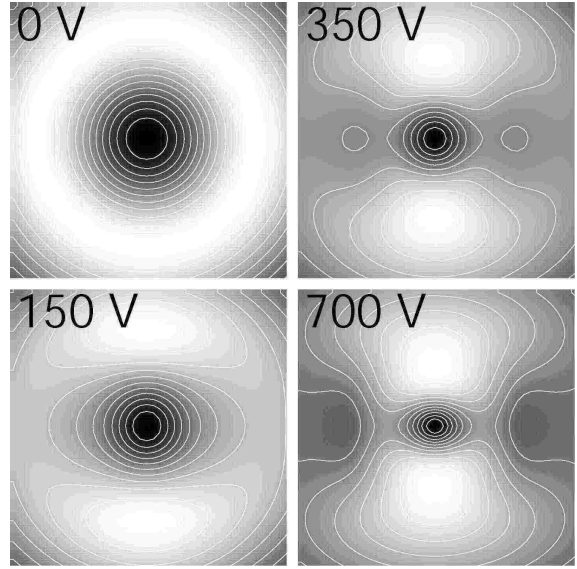}
\caption{Numerical results showing output spatial intensity distributions
of a vortex pair for different applied voltages.}
\label{fig.2}
\end{minipage}
\end{figure}

The above stability of the vortex pair is due to its phase structure. To prove this 
point we have carried out numerical calculations for the input field consisting 
of two temporal features analogous to Eqs. (\ref{pair}) and (\ref{fields}) 
but without azimuthal phase dependence
%
%%%%%%%%%%%%%%%%%%%novort%%%%%%%%%%%%%%%
\begin{equation}
B_{\pm}(x=0,\vec{r}) = \sqrt{I_{in}} r \exp(-r^{2}/d_{G}^{2})
\label{novort}
\end{equation}  
The field (\ref{novort}) has an input intensity distribution that 
is identical to the above discussed cases but lacks the topological 
phase structure present in the case of Eqs. (\ref{single}) and (\ref{pair}).  
Figure 3 shows results of the calculations. All parameters 
are the same as in Figs. 1 and 2.  In sharp contrast to Figs. 1 and 2 
the output distribution of the field does not contain any intensity zeros 
because of the diffraction. The output intensity for zero applied voltage is a 
bright ring with a smaller local maximum in the center that is about two times 
weaker than the ring. Increasing nonlinearity results in the transfer of energy 
from the ring to the center of the beam and the appearance of two intensity 
minima on the right and left. The intensity in these minima at the highest value 
of the applied voltage ($700~\rm V$) is about 0.2 of the maximum intensity 
in the center.

%
%Fig.3
%
\begin{figure}[!t]
\centering
\begin{minipage}[c]{7.5cm}
\includegraphics[width=7.5cm]{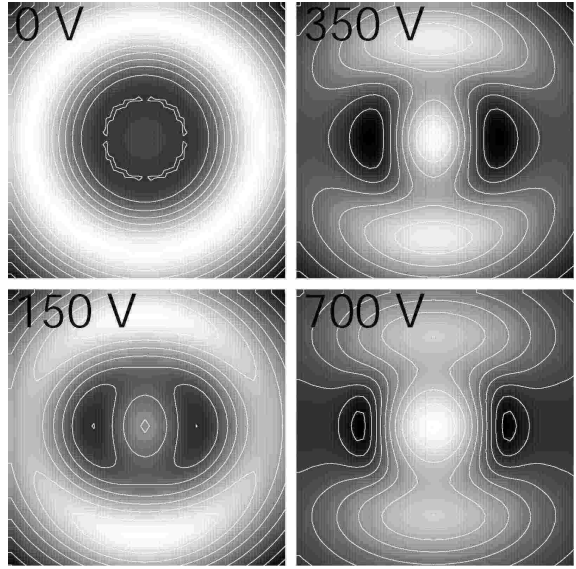}
\caption{Numerical results showing output spatial intensity distributions
of a hole in the intensity without phase structure. }
\label{fig.3}
\end{minipage}
\end{figure}

%
%Fig. 4
%
\begin{figure}[!t]
\centering
\begin{minipage}[c]{7.5cm}
\includegraphics[width=7.5cm]{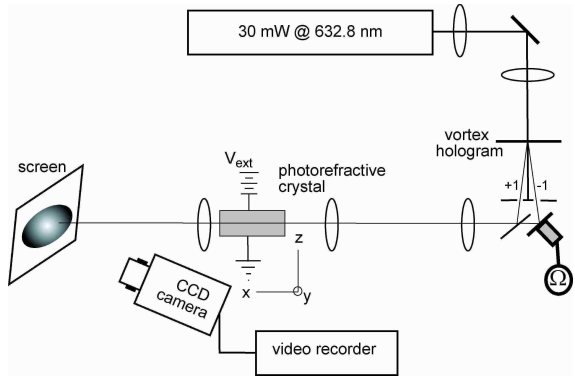}
\caption{Experimental setup.  }
\label{fig.4}
\end{minipage}
\end{figure}

\section{Experiment}
\label{experiment}
The experimental setup is shown in Fig. \ref{fig.4}.
A 30 mW He-Ne laser beam ($\lambda=0.63 ~\mu\rm m$) was passed through 
a variable beam splitter, and a system of lenses controlling the size of the beam waist.
Diffraction of the beam by a diffractive element with a fringe dislocation 
created  unit charge vortices 
of opposite signs in the first order diffracted beams to either side of the transmitted beam. 
One of the beams was phase modulated by reflection from a mirror mounted on a piezo-electric 
transducer driven by a sawtooth voltage, such that $\Omega\tau\gg 1.$  
The beams were then recombined and 
directed into a photorefractive crystal of SBN:60 doped with 0.002\% by
 weight Ce. The beams propagated perpendicular to the crystal $\hat c$-axis ($=z$ axis), 
and were polarized along it. The crystal measured $19~\rm mm$ along the direction of propagation, 
and was $5~\rm  mm$ wide along the  $\hat c$-axis. The experimentally measured value of the 
relevant component of the electrooptic tensor was found to be equal to $r_{33} = 130 ~\rm pm/V$.  
A variable dc voltage was applied 
along the $\hat c$-axis to control the value of the nonlinearity coefficient. The crystal 
was illuminated by a source of incoherent white light to increase the level of the 
effective saturation intensity.
Images of the beams at the output face of the crystal were recorded with a CCD camera. 
%
%Fig.5
%
\begin{figure}[!t]
\centering
\begin{minipage}[c]{7.5cm}
\includegraphics[width=7.5cm]{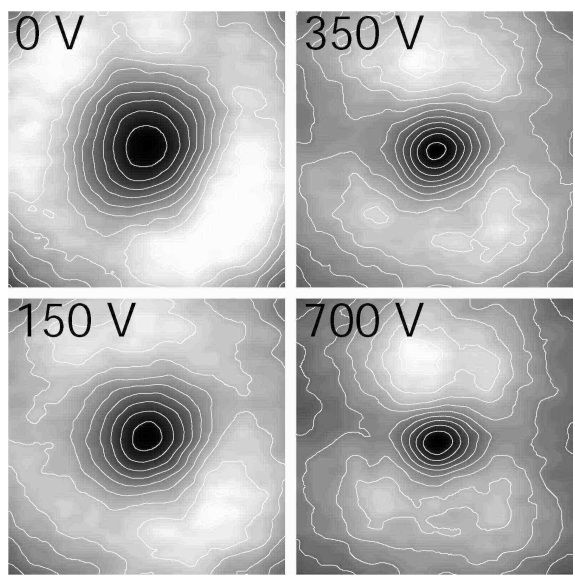}
\caption{Observed evolution of two spatially-coinciding counterrotating 
vortices for applied voltages of $0$, $150$, $350,$ and $700$ V. }
\label{fig.5}
\end{minipage}
\end{figure}

Figure \ref{fig.5} presents experimental output intensity distributions of the light beam with 
two embedded overlapping mutually incoherent vortices for different values of the 
applied voltage. The numbers on the frames give the applied  voltage in volts. 
The size of the frames is about $200~\mu$m.
The output intensity distribution in the frame with no applied voltage
corresponds to an annular ring with internal and external diameters of about $80$ 
and $260$ $\mu$m, respectively. 
Increasing the voltage results in the focusing of the vortex core and its stretching along 
the direction of the anisotropy which is clearly seen on the last frame corresponding to the 
$700$~V applied voltage. This stretching, however, is much smaller than that for a single 
vortex for similar parameters.  Experimental data on the distortions of a 
single vortex are given in Ref. \cite{vrpair}. 
Comparison with the results of the theoretical description shows good 
quantitative agreement between the experiment and the theory regarding 
the degree of ellipticity of the vortex core. Positions of intensity lobes (maxima) on 
the Gaussian beam above and below the vortex core in the
theoretical Fig. 3 also are in agreement with those on the experimental Fig. 5.  

\section{Conclusions}
\label{conclusions}
In summary we presented an experimental and theoretical study of the propagation of a
mutually incoherent pair of vortices. Incoherent coupling between the oppositely rotating vortices creates 
a symmetric attractive potential that arrests the rapid spreading and decay of an isolated vortex. 

M. S. was supported by the A. P. Sloan Foundation. 
We are grateful to E. Eilertsen and 
E. Rasmussen at Ris{\o} National Laboratory for preparation of the holographic mask.


\section*{References}
\begin{thebibliography}{99}
%
%
\bibitem{GinPit}
 Ginzburg V L and  Pitaevskii L P 1958  {\it Zh. Eksp. Teor. Fiz.} {\bf34} 1240 
 [1958 {\it Sov. Phys. JETP}
{\bf34} 858 ].
%
\bibitem{ZakRub}
Zakharov V E and Rubenchik  A M 1973
{\it Zh. Eksp. Teor. Fiz.} {\bf65} 997 
 [1974 {\it Sov. Phys. JETP} {\bf38} 494]. 
%
\bibitem{stripe_exp}
Mamaev A V,  Saffman M and Zozulya A A 
1996 {\it Phys. Rev. Lett.} {\bf 76} 2262 .
\bibitem{stripe_exp2}
Tikhonenko V, Christou J,  Luther-Davies B and  Kivshar Y S 1996
{\it Opt. Lett.} {\bf 21} 1129 .
%%
\bibitem{swa}
 Swartzlander G A, Jr. and  Law C T 1992 {\it Phys. Rev. Lett.} {\bf 69} 2503.
%
\bibitem{nlovortex}
 Coullet P,  Gil L, and  Rocca F 1989 {\it Opt. Commun.} {\bf73} 403;
 McDonald G S,  Syed K S and  Firth W J 1992 {\it Opt. Commun.} {\bf 94} 469; 
 Weiss C O 1992 {\it Phys. Rep.} {\bf219} 311;
Yu S. Kivshar 1993 {\it IEEE J. Quantum Electron.} {\bf 28} 250 ; 
K Staliunas 1994 {\it Chaos, Solitons \& Fractals} {\bf 4} 1783 ;
 Ackemann T, Kriege E and Lange W 1995 {\it Opt. Commun.} {\bf115}, 339.
%
\bibitem{atomicvortex}
 Lundh E, Pethick C J and  Smith H 1998 {\it Phys. Rev. }A {\bf 58} 4816 ;
 Denschlag J et al. 2000 {\it Science} {\bf 287} 97  ;
 Madison K W et al. 2000 {\it Phys. Rev. Lett.} {\bf 84} 806 ;
 Anderson B P et al. 2001 {\it Phys. Rev. Lett.} {\bf 86} 2926;
Leanhardt A E et al. 2003 {\it Phys. Rev. Lett.} {\bf 90} 140403 .

\bibitem{vrpair}
 Mamaev A V,  Saffman M and  Zozulya A A 1996 {\it  Phys. Rev. Lett.} {\bf 77} 4544.
%
\bibitem{nocircular}
Saffman M and  Zozulya A A
1998 {\it Opt. Lett.} {\bf 23} 1579 . 
%
\bibitem{vrtransient}
 Mamaev A V, Saffman M and Zozulya A A 
1997 {\it Phys. Rev. }A {\bf 56} R1713 . 
%
\bibitem{ZAPRA}
Zozulya A A and  Anderson D Z 1995 {\it Phys. Rev. }A {\bf{51}} 1520 .
%
\bibitem{ara96}
 Aranson I and  Steinberg V 1996 {\it Phys. Rev. }B {\bf53} 75 .
%
\bibitem{vrdecay}
 Mamaev A V,  Saffman M and Zozulya A A 1997 
{\it Phys. Rev. Lett.} {\bf 78} 2108.
%
\bibitem{lamb}
Lamb H 1932 {\it Hydrodynamics, sixth ed. } (Cambridge Univ. Press, Cambridge).
%
\bibitem{toys}
 Saffman M,  Benkert C and  Anderson D Z 1991 {\it Opt. Lett.} {\bf 16} 1993 ; 
 Saffman M, Montgomery D, Zozulya A A and Anderson D Z 1994 {\it Chaos, Solitons \& Fractals} {\bf 4} 2077 .
%
\bibitem{incohsol}
 Chen Z, Segev M, Coskun, T H and Christodoulides D N 1996 {\it  Opt. Lett.} {\bf21} 1436; 
 Chen Z et al. 1996 {\it Opt. Lett.} {\bf 21} 1821.
%
\bibitem{msazpra}
 Mamaev A V, Saffman M, Anderson D Z and Zozulya A A 1996 {\it Phys. Rev. }A {\bf 54} 870.
%

\end{thebibliography}
\end{document}